\pdfoutput=1
\documentclass[conference]{IEEEtran}
\IEEEoverridecommandlockouts
\usepackage[numbers]{natbib}
\usepackage{booktabs}
\usepackage{amsmath,amssymb,amsfonts}
\DeclareMathOperator*{\argmax}{arg\,max}

\usepackage{algorithmic}
\usepackage{graphicx}
\usepackage{textcomp}
\usepackage{xcolor}

\begin{document}
\title{Dimensionality reduction for prediction: \\Application to Bitcoin and Ethereum}

\author{\IEEEauthorblockN{Hugo Inzirillo}
\IEEEauthorblockA{\textit{CREST Institut Polytechnique de Paris} \\
\textit{Napoleon Group}\\
hugo.inzirillo@napoleon-group.com}
\and
\IEEEauthorblockN{Benjamin Mat}
\IEEEauthorblockA{\textit{Napoleon Group}\\
benjamin@napoleon-group.com}
}

\maketitle

\begin{abstract}

 \noindent The objective of this paper is to assess the performances of dimensionality reduction techniques to establish a link between cryptocurrencies. We have focused our analysis on the two most traded cryptocurrencies: Bitcoin and Ethereum. To perform our analysis, we took log returns and added some covariates to build our dataset. We first introduced the pearson correlation coefficient in order to have a preliminary assessment of the link between Bitcoin and Ethereum. We then reduced the dimension of our dataset using canonical correlation analysis and principal component analysis. After performing an analysis of the links between Bitcoin and Ethereum with both statistical techniques, we measured their performance on forecasting Ethereum returns with Bitcoin’s features. 

\end{abstract}
\section{Introduction}
The 21st century is characterised by the emergence of new technologies and the digitization of the financial industry. The digital age is still in its infancy, however since the financial crisis of 2008, in response to the weakness of the financial system, we have witnessed the birth and the multiplication of digital currencies. Since the last decade, 3000 new born cryptocurrencies were created according to \citet{ccy_types}. To categorize these new assets, \citet{ccy_types} distinguished three different type of cryptocurrencies : Bitcoin, Alcoins and Tokens. Bitcoin could be defined as a peer-to-peer electronic payment with no regulated counter-party in the transaction process, while Altcoin has been introduced as an alternative to Bitcoin. Ethereum (ETH), Ripple (XRP), LiteCoin (LTC) and Bitcoin Cash (BCH) are the largest altcoins (in term of market capitalization, source: CoinMarketCap \footnote{CoinMarketCap https://coinmarketcap.com/en/.}). Tokens  however are digital assets issued and tradable on blockchain. \\
This growing enthusiasm contradicts  the inherent nature of such an asset and its similarity with "classic" currencies. According to the standard definition:\textit{"Money serves three basic functions. By definition, it is a medium of exchange. It also serves as a unit of account and as a store of value"}, University of \citet{book_principle_economics}. In these days, possibilities to pay using Bitcoin are limited. Bitcoin are mainly stored in wallets and somewhat can appear as a store of value. Bitcoin could also be seen as an interesting tools for portfolio management, \textit{"as this asset is for the moment uncorrelated to other main asset classes, it becomes a very interesting diversifying asset"} \citet{bitcoin_portfolio_dartois}. Nevertheless, one Stylized Fact of digital currencies is that it a high volatility. The exponential trend in the price of digital assets makes them dedicated to speculation as mentionned by \citet{btc_market_facts}. Thus, the question could be: are they disruptive currencies created in response to state driven monetary policies or a new asset class born in the digital era.\\

In our empirical study we will consider two major cryptocurrencies (in term of market capitalization).  According CoinMarketCap \footnote{CoinMarketCap Dominance https://coinmarketcap.com/fr/charts/dominance-percentage}, Bitcoin (BTC) and Ethereum (ETH) respectively represent 62.41\% and 16.51\% of the crypto markets as of early January. Despite the fact that both assets rely on different technologies, we will try to establish a stable relation between them. To make it possible, we have introduced additional features from available market data. We assume that ETH is a dependant variable of the BTC. This assumption is made due to the fact that Bitcoin has the largest market capitalization. Our study will focus on multivariate analysis and dimensionality reduction with the use of Canonical Correlation Analysis (CCA),  a statistical technique introduced by \citet{h_hotteling}, and Principal Components Analysis, first introduced by \citet{k_pearson}. After performing dimensionnality reduction, we will look at forecasting accuracy to assess the effectiveness of  such  a technique. 
\section{Background}
\subsection{Principal Component Analysis (PCA)}
Principal Component Analysis (PCA), first introduced by \citet{k_pearson}, is a well known statistical technique for dimensionality reduction. The objective is to reduce the number of variables needed to explain the information contained in a dataset by finding new components that are combinations of the old variables. These new variables are linear maps of those in the original dataset.  This newly set of variable are orthogonal and the features are created to successively maximize variance.  Thus we can represent the PCA as a tool that determines the net effect of each variable on the total variance of the dataset and extracts the maximum variance possible from it by creating a new set of features. Principal components are obtained by solving an  eigenvalue/eigenvector problem.
\\
Let denote X the matrix of the standardized dataset, $X = (X^{(1)},...,X^{(q)})$, with $X^{(i)} \in \mathbb{R}^{n}$ .Let denote $\Omega_{X}$ the covariance matrix of the dataset such as
\begin{equation}
    \Omega_{X} = X' X =\begin{pmatrix}
   \sigma_{1}^{2} & \ldots & \sigma_{1,q} \\
   \vdots & \ddots & \vdots \\
   \sigma_{q,1}  & \ldots & \sigma_{q}^{2} \\
\end{pmatrix}
\end{equation}

In order to retrieve the principal components of our dataset, we need to compute the eigenvectors of $\Omega_{X}$.  
Let us denote $\Lambda$, the diagonal matrix of eigenvalues and $V$ the matrix of eigenvector. We have: 

\begin{equation}
    \Omega_{X} V = \Lambda V \Leftrightarrow (\Omega_{X}-\Lambda I_{q}) V = 0
\end{equation}

We then compute the characteristic equation: 
\begin{equation}
     \left|\Omega_{X}-\Lambda I_{q} \right| = 0
\end{equation}

The values of eigenvalues can be found by solving $det(\Omega_{X} - \Lambda I_{q}) = 0$; the values of the eigenvectors are then computed by solving $(\Omega_{X} V) = \Lambda$
\\
\\
The eigenvectors of $\Omega_{X}$ are the directions of the axes where there is the most variance, eigenvalues being the coefficients attached to eigenvectors. By ranking the eigenvectors in order of their eigenvalues, highest to lowest, we get the principal components in order of significance. Thus the new dataset, noted W, composed of the Principal Components is equal to: 
\begin{equation}
    W = X * V 
\end{equation}
\\
In order to perform dimensionality reduction, we then need to choose how many principal components we need to keep; i.e remove principal components with low eigenvalues. Once we set the number of principal component to keep, we remove columns from W corresponding to principal components we want to drop to obtain the new dataset which has a lower dimension than the original dataset. 
\subsection{Canonical Correlation Analysis (CCA)}
\begin{figure}
    \centering
    \includegraphics[scale=0.45]{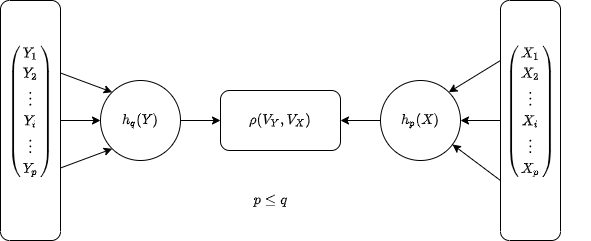}
    \caption{CCA Schema}
    \label{fig:cca_schema}
\end{figure}

{Another technique used for dimensionality reduction is the Canonical Correlation Analysis (CCA) first introduced by \citet{h_hotteling}. The objective of this multivariate technique is to measure the cross relationship between random variables. The building process of CCA can be represented by the Figure \ref{fig:cca_schema}.}
\\
\\
Let define two datasets: $Y=(Y^{(1)},...,Y^{(p)})$ and $X=(X^{(1)},...,X^{(q)})$, with $Y^{(i)} \in \mathbb{R}^{n}$ and $X^{(j)} \in \mathbb{R}^{n}$ . 
Let also define the canonical variate as $V_{Y}, V_{X} \in \mathcal{L}(\mathbb{R}^{n},\mathbb{R}^{n})$.\\ \\
The canonical correlation problem consist in the maximisation of correlation coefficient:

\begin{equation}
 \max \rho(V_{Y}, V_{X})  
\end{equation}
\\
With $\rho_{V_{Y},V_{X}}$ the pearson correlation metric equal to $\frac{V_{Y}V_{X}}{ {||V_{Y}||}_2 {||V_{X}||}_2}$. We can set $V_{Y}$ and $V_{X}$ :

\begin{equation}
        V_{Y}:= h_{p}(Y) = \sum_{i=1}^{p} \alpha_{i}Y^{i}  \quad\textrm{and}\quad  V_{X}:= h_{q}(X) = \sum_{j=1}^{q} \beta_{j}X^{j} 
\end{equation}
hence we can rewrite with $\alpha \in \mathbb{R}^{p} $ and $\beta \in \mathbb{R}^{q} $:
\begin{equation}
    \label{eq:initial_max_program}
  (\alpha^{*},\beta^{*}) = {\argmax_{\alpha,\beta} \frac{\alpha^{T} Y^{T}X \beta}{\sqrt{\alpha^{T} Y^{T}Y\alpha \beta^{T} X^{T}X\beta}}}  
\end{equation}

\begin{equation}
  (\alpha^{*},\beta^{*}) = {\argmax_{\alpha,\beta} \frac{\alpha^{T} \Omega_{YX} \beta}{\underbrace{\sqrt{\alpha^{T} \Omega_{YY}\alpha}}_{{||A||}_2}\underbrace{\sqrt{\beta^{T} \Omega_{XX}\beta}}_{{||B||}_2}}}  
\end{equation}
\\
Replacing ${||A||}_{2}^{2}=\alpha^{T} \Omega_{YY}\alpha$ and  ${||B||}_{2}^{2}=\beta^{T} \Omega_{XX}\beta$ we obtain:

\begin{equation}
        \label{eq:opt_with_constraints}
  (\alpha^{*},\beta^{*}) = {\argmax_{\alpha,\beta} \frac{\alpha^{T} \Omega_{YX} \beta}{{||A||}_{2}{||B||}_{2}}} 
  \quad\textrm{s.t}\quad {||A||}_{2}^{2}=1  \quad {||B||}_{2}^{2}=1
\end{equation}

On the geometrical point of view we can rewrite the problem using \eqref{eq:initial_max_program} and constraints defined in \eqref{eq:opt_with_constraints}  by transforming the numerator. We know that $\Omega_{YX} = Y^{T}X$ so we can write:

\begin{equation}
        \alpha \Omega_{YX} \beta  =\left(Y\alpha\right) ^{T} X \beta = A ^{T} B
\end{equation}

hence we can rewrite:

\begin{equation}
\label{eq:max_corr}
  (A^{*},B^{*}) = {\argmax_{A,B} \frac{A^{T}B}{{||A||}_2{||B||}_2}}
\end{equation}

\begin{figure}
    \centering
    \includegraphics[scale=0.8]{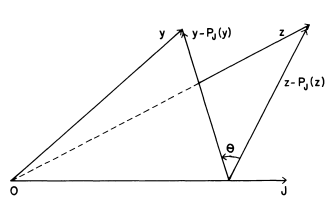}
    \caption{The simple correlation coefficient \citet{geo_corr}}
    \label{fig:corr_geom}
 \end{figure}

When we look at the problem in this form we can  see what happens geometrically. Here the goal is to maximise the correlation which is the  equivalentof  minimising the angle $\theta$  between A and B, our two canonical variables, represented in figure \ref{fig:corr_geom}.
\begin{equation}
    \rho_{A,B} = \cos(\theta) = \frac{A^{T}B}{{||A||}_2{||B||}_2}
\end{equation}

\section{Study}
\subsection{Data}
{For our study we downloaded Bitcoin \& Ethereum high, low, open and close prices as well as traded volumes. We retrived historical data using Yahoo Finance. Our data sample starts from 2017-12-15 to 2020-12-15}. Other more specialized data sources could have been used for this purpose, but we chose Yahoo Finance to make the analysis as easily replicable as possible. All time series are standardized before data processing. 

\subsection{Methodology}

Toward assessing potential existing correlations between Bitcoin and Ethereum, we added the following features to the original dataset: log returns of the closing price, high minus low $(HML)$ Factor, a volatility momentum factor $(VMOM)$ and an intraday variance factor denoted $(IV)$.\\
\\
Log returns are computed using the following formula where $P_{t}^{i}$ denote the Price of the asset at time $t \in \{1,...,n\} $ for asset $i \in \{1,2\} $:
$$
    R_{t}^{i} := \log\left(P_{t}^{i}\right) - \left(P_{t-1}^{i}\right) 
$$
\\
The $HML$ factor is given:
$$
    HML_{t}^{i} := H_{t}^{i}-L_{t}^{i} 
$$
\\
The $(VMOM)$ factor  is built on weekly basis : 
\begin{equation}
    VMOM_{t}^{i} =\frac{P_{t}^{i}-P_{t-7}^{i}}{\frac{1}{22}\sum_{\tau=t-22}^{t}HML_{\tau}^{i}}
\end{equation}
with $t \in \{23,...,n\} $ \\

The proxy of the intraday volatility factor is computed on the assuption of squared returns. We can define $\theta^{i} \in R^{4}_{+}$. $\theta:= \{O,H,L,C\}$.The proxy of intraday volatility of the i-th asset is given by:

\begin{equation}
    f_{t}(\theta^{i}):= \psi_{O,C}(H,L)
\end{equation}

With 

$$
    \psi_{O,C}=\begin{cases}
    \log \left(\frac{H}{L}\right), & \text{if $O\geq C$}.\\
    \log \left(\frac{L}{H}\right), & \text{if $O<C$}.
  \end{cases}
$$

Using the following formulae under the assumption $\mathbb{E}\left[ f_{t}(\theta^{i}) \right]=0$ we then have $\mathbb{V}\left[ f_{t}(\theta^{i}) \right]=\left[\mathbb{E}\left[ f_{t}(\theta^{i})^2\right]\right]$. We can then compute the our Intraday Variance Factor:
\begin{equation}
    IV_{t}^{i} = f_{t}(\theta^{i})^2
\end{equation}

Once we obtained our standardized vectors we removed features which have a Pearson’s correlation with another feature greater than 99.5\% .
In our application of principal components analysis, we will assess Bitcoin’s features of the dataset. To carry out this analysis we used the PCA package provided by sklearn. 
In our application of Canonical Correlation Analysis, we will assess log returns of Ethereum prices, Ethereum Volume and Ethereum proxy for intraday volatility versus Bitcoin’s features available in the datasets. We can now define our dataset $D= \{(Y_1,X_1),...,(Y_n,X_n)\}$ with $Y \in\mathbb{R}^{n\times p} $  and $X \in \mathbb{R}^{n\times q}$ with $p \leq q$. To carry out this analysis we used the package CanCorr provided by statmodels in python.
Before looking to dimensionnality reduction results, we can visualize the selected features (daily return, closing price, volume, HML, VMoM, IV) for both Bitcoin and Ethereum.

\begin{figure}
\centering
\includegraphics[scale=0.38]{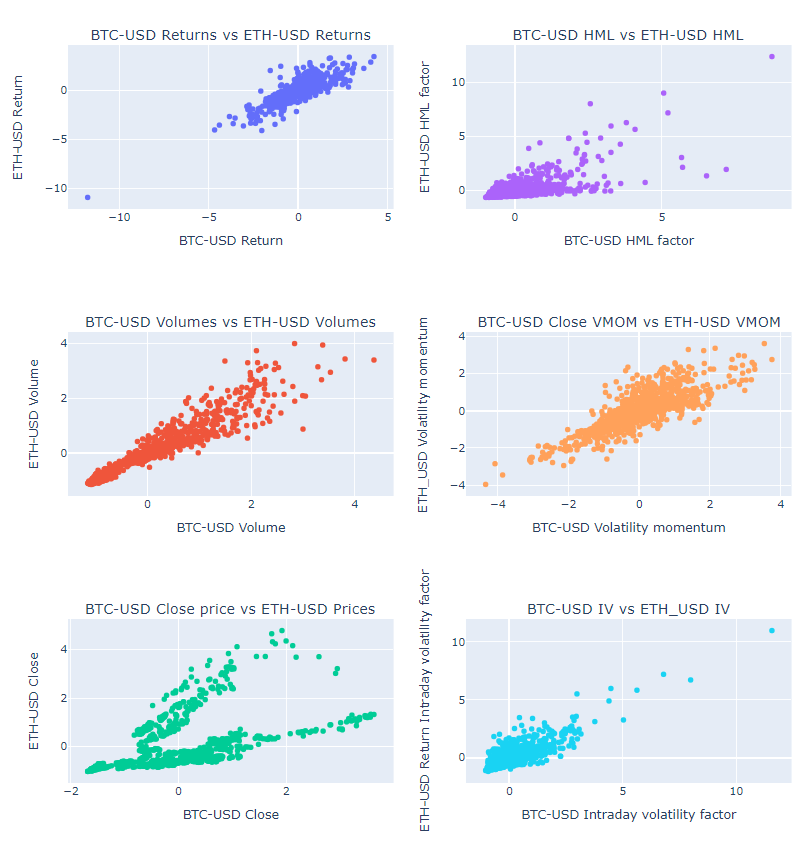}
\caption{Representation of the standardized features}
\label{fig:my_label}
\end{figure}

\subsection{Representation of the Pearson correlation in the dataset}
\begin{figure}
    \centering
    \includegraphics[scale=0.40]{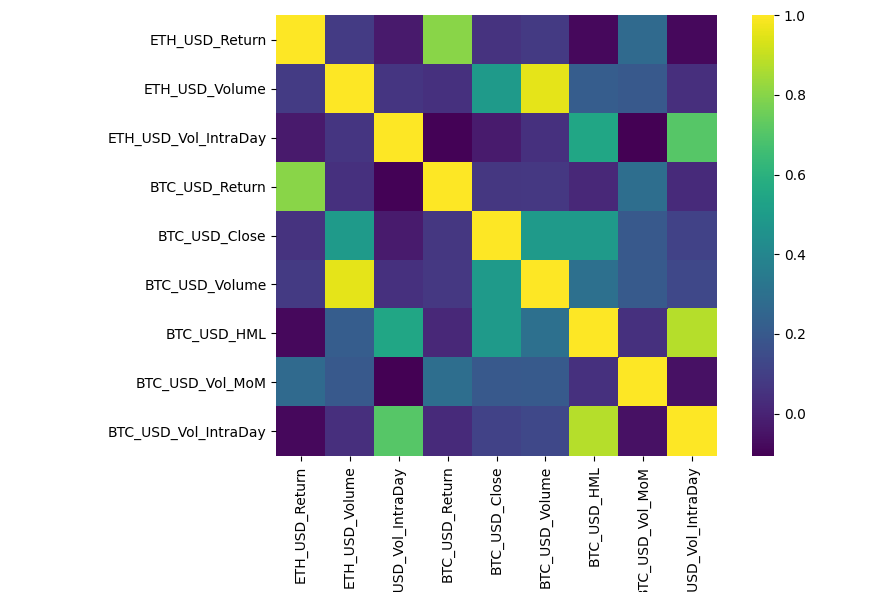}
    \caption{Correlation matrix of the dataset}
\label{fig:pearson_correl}
\end{figure}

{Pearson's correlations between Bitcoin and Ethereum features are presented in Figure \ref{fig:pearson_correl}.}

We denote four significant Pearson's correlation. These correlation are: between Bitcoin and the Ethereum Volume ($\rho_{BTC_{Volume},ETH_{Volume}}=0.95 $), Bitcoin volume and Bitcoin HML factor ($\rho_{BTC_{Volume},ETH_{HML}}=0.88 $), Bitcoin and the Ethereum daily return ($\rho_{BTC_{Return},ETH_{Return}}=0.8 $), Bitcoin Intraday volatility and the Ethereum intraday volatility ($\rho_{BTC_{IV},ETH_{IV}}=0.71 $).

\subsection{Principal Components Analysis}

In order to perform the principal components analysis we take the following features to build our initial dataset: Bitcoin's closing prince, Bitcoin's daily return, Bitcoin's daily volume, Bitcoin's HML factor, Bitcoin Volatility Momentum factor and the Bitcoin intraday volatility factor. The result of the principal components analysis are presented in table I.

\begin{table}
  \centering
	\begin{tabular}{|c|c|c|c|}
	
	\hline
	Components &       Eigenvalues &       \% of Variance &       Cumulative \% \\
	\midrule
	\hline
	Component 1    &  1.4101 &  37.19\% & 37.19\% \\
    Component 2     &  0.8997 & 23.73\%  & 60.09\% \\
    Component 3      &  0.63073 & 16.63\%  & 77.55\%  \\
    Component 4      & 0.45462 &  11.99\% & 89.45\%  \\
    Component 5      &  0.3800 &  10.02\% & 99.56\% \\
    Component 6 &  0.01651 &  0.44\% & 100\% \\
    \hline

\end{tabular}
\caption{Principle Component Analysis Total Variance Explained.}
\end{table}

\begin{figure}
\centering
\includegraphics[scale=0.18]{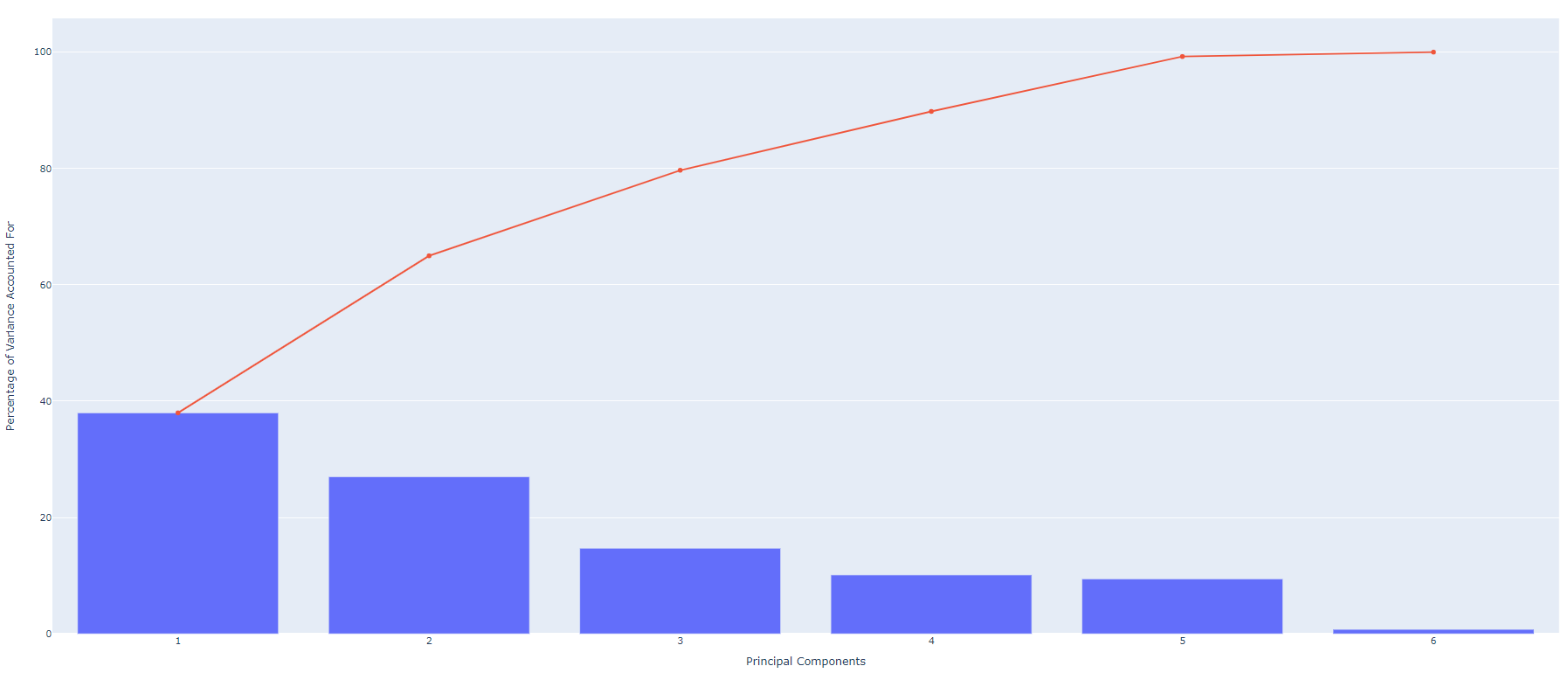}
\caption{Representation of principal components}
\label{fig:my_label}
\end{figure}

We can notice that none of the principal components capture a percentage of variance of the intial dataset higher than 50\%. Indeed, only the sixth principal components is not significant with 0.44\% of the variance of the initial dataset explained. We now look at the covariance between the initial Bitcoin's features and the pincipal components obtained.

\begin{table}
\resizebox{\columnwidth}{!}{
    \centering
    \begin{tabular}{|c|c|c|c|c|c|c|}
    
    \hline
    {} & First PC & Second  PC & Third PC & Fourth PC & Fifth PC & Sixth PC \\
    \midrule
    \hline
    BTC\_USD\_Close        &  0.7780 &  0.1444 & -0.2130 &  0.0311 & -0.5710 &  0.0382 \\
    BTC\_USD\_Return       &  0.2521 & -0.4847 &  0.4099 & -0.7247 & -0.0911 & -0.0044 \\
    BTC\_USD\_Volume       &  0.7906 &  0.0720 & -0.4380 & -0.1718 &  0.3852 & -0.0022 \\
    BTC\_USD\_HML          &  0.6451 &  0.5475 &  0.4979 &  0.1194 & -0.0334 & -0.1442 \\
    BTC\_USD\_Vol\_MoM      &  0.4881 & -0.7536 &  0.2061 &  0.3820 &  0.0736 &  0.0024 \\
    BTC\_USD\_Vol\_IntraDay &  0.3886 &  0.5678 &  0.6704 &  0.0687 &  0.2409 &  0.1201 \\
    \hline
    
    \end{tabular}}
    \caption{Covariance between Principal Component (PC) and the initial dataset.}
\end{table}

Bitcoin's closing price and daily volume are the  most correlated features to the first principal component with a respective correlation of 0.78 and 0.79. We can also notice that Bitcoin HML factor  has a correlation to the first principal component of 0.65. Regarding the second principal component we notice a significant negative correlation with the Bitcoin volatility momentum factor of -0.75. We also note weaker correlation with Bitcoin HML factor and Bitcoin $IV$ factor with a respective correlation of 0.55 and 0.57. The third principal component is most correlated with the Bitcoin $IV$ factor with a correlation of 0.67. The fourth principal component registers a significant negative correlation with Bitcoin daily returns of -0.72. The fifth and sixth principal components do not show significant correlation.

\subsection{Canonical Correlation Analysis}
The results of the Canonical Correlation Analysis are presented in Table III. 

\begin{table}
    \centering

\resizebox{\columnwidth}{!}{

\begin{tabular}{|c|c|c|c|c|c|}

\hline
Canonical Correlation & Wilks' lambda & Num DF &   Den DF &  F Value &        Pr > F \\
\hline
\midrule
             0.965987 &    0.00977338 &     18 &  2701.63 &  620.778 &             0 \\
             0.855689 &      0.146156 &     10 &     1912 &  308.926 &             0 \\
             0.673962 &      0.545776 &      4 &      957 &  199.117 &  3.16494e-124 \\
\hline

\end{tabular}}
    \caption{Canonical Correlation Results}
    \label{tab:canncorr}
\end{table}
First, we test the null hypothesis that all $p$ canonical variate pairs are uncorrelated. Recall of the null hypothesis: 
$$
    H_{0}= \rho_{1}^{*}=\rho_{i}^{*}=...=\rho_{p}^{*} = 0 
$$

with $\rho_{i}^{*}$ for $i \in \{1,..,p\}$ the canonical correlations. In our case $p=3$ 
\\
Wilks' Lambda for each correlation between the canonical pair $(V_{X_{i}},V_{Y_{i}}) $ for $i \in \{1,..,p\}$ are as follow: 
$\Lambda_{1} = 0.00977338$ with $\rho_{1}^{*} = 0$, $\Lambda_{2} = 0.146156$ with $\rho_{2}^{*} =0$, $\Lambda_{3} = 0.545776$ with $\rho_{3}^{*} <0.0001$. Thus we can reject the null hypothesis and conclude that $\rho_{1}^{*} \neq 0$,  $\rho_{2}^{*} \neq 0$ and $\rho_{3}^{*} \neq 0$.
\\
We see that 96.60\% of the variation in $V_{X_{1}}$  is explained by the variation in $V_{Y_{1}}$, 85.57\% of the variation in $V_{X_{2}}$ is explained by $V_{Y_{2}}$ and 67.40\% of the variation in $V_{X_{3}}$ is explained by $V_{Y_{3}}$. Thus, our two dataset have a positive correlation.\\ 
\begin{figure}
    \centering
    \includegraphics[scale=0.40]{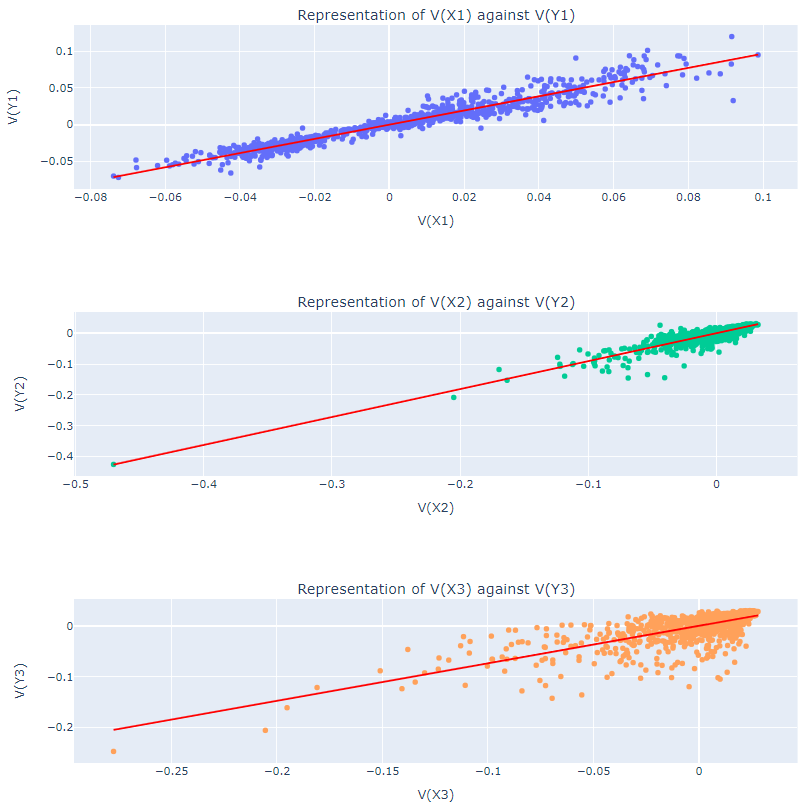}
    \caption{Representation of canonical pairs}
    \label{fig:my_label}
\end{figure}
The plot of the first canonical variate pair show a strong relationship between this canonical pair. The second canonical variate pair plot is more scattered, but we still have a robust relationship between this canonical variates. The third plot of the pair show  weaker relationship between the two canonical variates.





To interpret each canonical variate, we compute the correlations between each variable and the corresponding canonical variate. We thus have : 

\begin{table}
    \centering
	\begin{tabular}{|c|c|c|c|}
	
	\hline
	 &       $V_{X1}$ &       $V_{X2}$ &       $V_{X3}$ \\
	\midrule
	\hline
	BTC\_USD\_Close        &  0.511188 &  0.043281 & -0.048267 \\
    BTC\_USD\_Return       &  0.110806 & -0.838333 & -0.526131 \\
    BTC\_USD\_Volume       &  0.966602 &  0.162986 & -0.191605 \\
    BTC\_USD\_HML          &  0.096883 &  0.444519 & -0.640531 \\
    BTC\_USD\_Vol\_MoM      &  0.238879 & -0.283286 & -0.100641 \\
    BTC\_USD\_Vol\_IntraDay & -0.114520 &  0.496521 & -0.8293 \\
    \hline
    
\end{tabular}
\caption{Correlation between Bitcoin related canonical variates and Bitcoin features}
\end{table}
Looking at the first canonical variable for Bitcoin, we see that this variable is strongly correlated with the Bitcoin’s volume. Moreover we also notice a weaker correlation with the Bitcoin’s closing prices. Therefore, the variation of this canonical variate is mostly driven by these two features. For the second canonical variable related to Bitcoin, we see significant negative correlation with the Bitcoin’s return and a  weaker positive correlation with the Bitcoin's HML factor and the Bitcoin’s IV factor. The third canonical variate has significant negative correlation with the Bitcoin’s IV factor and  weaker negative correlations with the Bitcoin's HML factor and Bitcoin's return. Moreover this canonical variate is negatively correlated with all the Bitcoin features in our dataset.

\begin{table}
    \centering
	\begin{tabular}{|c|c|c|c|}
	\hline
	
	 &       $V_{Y1}$ &       $V_{Y2}$ &       $V_{Y3}$ \\
	\midrule
	\hline
	ETH\_USD\_Return       &  0.140567 & -0.841493 & -0.521661 \\
    ETH\_USD\_Volume       &  0.976243 &  0.154453 & -0.151970 \\
    ETH\_USD\_Vol\_IntraDay & -0.145664 &  0.531467 & -0.834461 \\
	\hline
	
	\end{tabular}
\caption{Correlation between Ethereum related canonical variates and Ethereum features}
\end{table}
Looking at the first canonical variable for Ethereum, we see that this variable is strongly correlated with the Ethereum’s volume.  Therefore, the variation of this canonical variate is mostly driven by these features as the Pearson’s correlation for the other feature are relatively small. For the second canonical variable related to Ethereum, we see significant negative correlation with the Ethereum’s Return factor and a weaker positive correlation with the Ethereum's IV factor. \\

The third canonical variate has significant negative correlation with the Ethereum’s IV factor and  weaker negative correlations with Etehreum's daily returns. Moreover this canonical variate is negatively correlated with all the Ethereum’s features in our dataset. The correlation between the canonical pair related to Ethereum and the Bitcoin’s feature are: \\

\begin{table}
    \centering
	\begin{tabular}{|c|c|c|c|}
	
	\hline
	 &       $V_{Y1}$ &       $V_{Y2}$ &       $V_{Y3}$ \\
	\midrule
	\hline
	ETH\_USD\_Return       &  0.135786 & -0.720057 & -0.351580 \\
    ETH\_USD\_Volume       &  0.943037 &  0.132164 & -0.102422 \\
    ETH\_USD\_Vol\_IntraDay & -0.140710 &  0.454771 & -0.562395 \\
	\hline
	
\end{tabular}
\caption{Correlation between Bitcoin related canonical variates and Ethereum’s features}
\end{table}
We can see a strong correlation between the first canonical feature related to Bitcoin and the Ethereum volume which is consistent to the fact that the first canonical correlation is very high. Similarly, we can see a strong negative correlation between the second canonical feature related to Bitcoin and the Ethereum's return which is consistent to the fact that the second canonical correlation is also high. The third canonical variate shows weaker and negative correlation with Ethereum's features which is consistent with the fact that the third canonical pair is less correlated than the previous one and has a negative correlation with Bitcoin's features. Nonetheless, the Ethereum's IV factor is the one which is the most negatively correlated with the third Bitcoin's related canonical variate. The correlation between the canonical pair related to Ethereum and the Bitcoin’s feature are:\\ 

\begin{table}
    \centering
	\begin{tabular}{|c|c|c|c|}
	\hline
	
	 &       $V_{Y1}$ &       $V_{Y2}$ &       $V_{Y3}$ \\
	\midrule
	\hline
	BTC\_USD\_Close        &  0.493801 &  0.037035 & -0.032530 \\
    BTC\_USD\_Return       &  0.107037 & -0.717353 & -0.354592 \\
    BTC\_USD\_Volume       &  0.933725 &  0.139465 & -0.129134 \\
    BTC\_USD\_HML          &  0.093587 &  0.380370 & -0.431693 \\
    BTC\_USD\_Vol\_MoM      &  0.230754 & -0.242405 & -0.067828 \\
    BTC\_USD\_Vol\_IntraDay & -0.110625 &  0.424867 & -0.558916 \\
	\hline
	
	\end{tabular}
\caption{Correlation between Ethereum related canonical variates and Bitcoin’s features}
\end{table}

We can see a strong correlation between the first canonical feature related to Ethereum and the Bitcoin volume which is consistent to the fact that the first canonical correlation is very high. Similarly, we can see a strong negative correlation between the second canonical feature related to Ethereum and the Bitcoin's return which is consistent to the fact that the second canonical correlation is also high. The third canonical variate shows weaker and has a negeative correlation with Bitcoin's features which is consistent with the fact that the third canonical pair is less correlated than the previous one and has a negative correlation with Etehreum's features. Nonetheless, the Bitcoin's IV factor is the one which is the most negatively correlated with the third Bitcoin's related canonical variate.\\
\subsection{Forecasting Ethereum returns with Bitcoin's features}
\subsubsection{Framework}
From the subset built with the PCA and the CCA we will now assess the performance and the level of significance for both techniques. In order to perfom the best evaluation of these statistical techniques, we will test the significance of the coefficent for a Linear Model. We will also use two additionnal measure of accuracy, the  residual mean squared error (RMSE) and the mean absolute error (MAE) to complete our analysis.
$$
            RMSE = \sqrt{\frac{\sum_{t=1}^{n} \left( Y_t- \hat{Y}_t \right)^2}{n}}  \quad
             MAE = \frac{\sum_{t=1}^{n} \left| Y_t- \hat{Y}_t \right|}{n}
$$
We will perform the linear regression on our dataset $X=(X^{(1)},...,X^{(j)},...,X^{(q)})$ and $X^{j} \in \mathbb{R}^{n}$. The $\hat{Y}$ will be the predicted estimated values of $Y_{i}$. $\Bar{Y_{i}}$ will be the predicted value according to ${X_{i}}$. 

\begin{equation}
    \hat{Y_t} = f\left(X,\beta \right)+\epsilon_t \quad\textrm{,}\quad \beta \in \mathbb{R}^{q}
\end{equation}

After the estimation of $\beta$ we will try to predict the return of the ETH according to BTC features. The results obtained will be compare to the result obtained with the initial dataset. 

\subsubsection{In sample Assessment of the ETH return based on BTC features}
Linear regression result for the initial dataset are presented below:
\begin{table}
 \centering
\begin{tabular}{|l|r|r|r|}
    
    \hline
    {} &    parameters &     t-stat &        p-value \\
    \midrule
    \hline
        $\beta_{BTC_{Close}}$ &  0.05565 &   1.61883 &  1.0581e-01 \\
        $\beta_{BTC_{Return}}$ &  0.81140 &  39.57402 & 1.6108e-203 \\
        $\beta_{BTC_{Volume}}$ &  0.02801 &   1.48609 &  1.3759e-01 \\
        $\beta_{BTC_{HML}}$ & -0.19462 &  -2.30157 &  2.1574e-02  \\
        $\beta_{BTC_{VMOM}}$ &  0.03041 &   1.74491 &  8.1323e-02 \\
        $\beta_{BTC_{IV}}$ &  0.04712 &   0.64437 &  5.1949e-01 \\
    
    \hline
\end{tabular}
\caption{Linear Regression results for the Bitcoin dataset.}
\end{table}
\\

By looking at the p-value we can see that only the $BTC_{Return}$ is significant with $pval(BTC_{Return}) < 0.01$. We can also noticed that the $BTC_{HML}$ feature is significant if we choose $\alpha=0.05$, as $pval(BTC_{HML}) < 0.05 $. Thus only 2 out of 6 features are statistically significant for $\alpha=0.05$. \\

Linear regression result for the dataset built from covariate pairs are presented below: \\
\begin{table}
    \centering
    \begin{tabular}{|l|r|r|r|}

\hline
 &     $parameters$ &     $t-stat$ &        $p-value$ \\
\midrule
\hline
$\beta_{ V_{X1}}$ &   3.304537 &   7.277908 &   7.055839e-13 \\

$\beta_{V_{X2}}$ & -17.251835 & -38.444331 &  2.387102e-196 \\

$\beta_{V_{X3}}$ &  -8.581909 & -19.095554 &   3.919702e-69 \\
\hline

\end{tabular}
    \caption{Linear regression result for covariate pairs}
    \label{tab:my_label}
\end{table}\\

Looking at the p-value for each covariate pairs we can see that the coefficient $\beta_{V_{X1}}$,$\beta_{V_{X2}}$ and $\beta_{V_{X3}}$ are significantly different from 0,  $ pval< 0.01$. 
Linear regression result for the dataset built with Principal Components and a 75\% variance contribution threshold:\\
\begin{table}
 \centering
\begin{tabular}{|l|r|r|r|}
    
    \hline
     &    $parameters$ &     $t-stat$ &        $p-value$ \\
    \midrule
    \hline
        $PC_{1}$ & 0.12605 &   7.20522 &  1.1729e-12 \\
        $PC_{2}$ & -0.38549 & -17.60144 &  2.6716e-60 \\
        $PC_{3}$ &  0.24091 &   9.21008 & 2.0032e-19 \\
    
    \hline
\end{tabular}
\caption{Linear Regression results for the first three principal components build with the Bitcoin dataset.}
\end{table}
Looking at the p-value for each principal components we can see that all the coefficient $\beta_{PC_{1}},\beta_{PC_{2}}$ and $\beta_{PC_{3}}$ are significantly different from 0 with $ pval< 0.01$.
Linear regression result for the dataset built with Principal Components and a 85\% variance contribution threshold:
 \\
\begin{table}
\centering
    \begin{tabular}{|l|r|r|r|}

\hline
{} &    params &     t-stat &        p-value \\
\midrule
\hline
$PC_{1}$ &  0.12605 &  10.10497 &   7.03285e-23 \\

$PC_{2}$ & -0.38549 & -24.68517 &  1.68542e-104 \\

$PC_{3}$ &  0.24091 &  12.91669 &   2.67366e-35 \\

$PC_{4}$ & -0.66896 & -30.45100 &  5.96977e-143 \\
\hline

\end{tabular}
\caption{Linear Regression results for the first fourth principal components build with the Bitcoin dataset.}
\end{table}

Looking at the p-value for each principal components we can see that all the coefficient $\beta_{PC_{1}},\beta_{PC_{2}}$, $\beta_{PC_{3}}$ and $\beta_{PC_{4}}$ are significantly different from 0 with $ pval< 0.01.$
The RMSE and the MAE for each model are presented below: 
\begin{table}
 \centering
\begin{tabular}{|l|l|r|}
    
    \hline
    Model & RMSE & MAE \\
    \midrule
    \hline
        initial dataset            & 0.45374 & 0.32480 \\
        CCA                         & 0.45377 & 0.32505 \\
        PCA with threshold 85\%     & 0.45799 & 0.32776 \\
        PCA with threshold 75\%     & 0.64264 & 0.49054 \\
    
    \hline
\end{tabular}
\caption{RMSE and MAE for each model.}
\label{tab:rmse_mae}
\end{table}
The model who minimize both RMSE and MAE is the initial dataset. This result is consistent with the fact that the best dataset is the one with the most information. Regarding the PCA and the CCA model, we see that the CCA model minimize the RMSE and the MAE compare to PCA models. However, the difference are small, i.e inferior to 0.01. 
\subsubsection{Out of sample Assessment of the ETH return based on BTC features}
The out of sample forecasting period start date is from 2020-12-16 to 2021-01-31. Due to the high volatility of the cryptocurrency market, we choose a relative small period for our out of sample assessment. By doing so, we aim to minimize the impact of exogenous shock on our assessment. The quality of the prediction will be assessed on the same criteria as the in sample assessment.
The RMSE and the MAE for each model are presented below: \\

\begin{table}
 \centering
\begin{tabular}{|l|l|r|}
    
    \hline
    Model & RMSE & MAE \\
    \midrule
    \hline
        initial dataset            & 0.8028 & 0.5955 \\
        CCA                         & 0.8031 & 0.5959 \\
        PCA with threshold 85\%     & 1.3274  & 1.0184 \\
        PCA with threshold 75\%     &1.4056  & 1.0179 \\
    
    \hline
\end{tabular}
\caption{RMSE and MAE for each model.}
\label{tab:rmse_mae_prediction}
\end{table}
The model who minimize both RMSE and MAE is still the initial dataset. Compare to in sample analysis, CCA techniques show better result than PCA models by minimizing both RMSE and MAE. Indeed, results obtained by CCA are closed to the one obtained by the intial dataset whereas PCA techniques record higher RMSE and MAE.  

\section{Conclusion}
 In this paper, we looked at the potential links between the most traded cryptocurrencies: Bitcoin and Ethereum. To do so we first looked at the Pearson's correlation between features, namely daily log returns, closing prices, traded volume, HML factor, volatility momentum factor and a proxy of the intraday volatility. We noticed four significant Pearson's correlation between features within the dataset. These correlation were between Bitcoin and Ethereum Volume,  Bitcoin and Ethereum daily return and  the Bitcoin and the Ethereum intraday volatility.\\

In order to represent the overall underlying links between these assets, we applied a Canonical Components Analysis. These dimensionality reduction techniques allowed us to summarized relationship between datasets that represent the behaviour of both assets while preserving the main aspects of this relationship. We noticed an existing relationship between the Bitcoin and the Etehreum as the Wilk's lambda test rejected the hypothesis of uncorrelated features between these two assets with an error threshold lower than 0.1\%. Moreover, the correlation between the Bitcoin and the Ethereum appeared robust as the two first canonical correlation were superior to 0.85.\\

The objective of this paper was also to compare dimensionality reduction methods applied to cryptocurrency market. We noticed that the Canonical Correlation Analysis is a performant tool to capture existing links between cryptocurrencies as it allows to efficiently represent the relationship between them, while reducing the numbers of features in the dataset.  We also find the CCA more convenient than the use of Pearson correlation and PCA to analyze relationship between Bitcoin and Ethereum. This statement is determined by the fact that the CCA allows us to connect our subsets without conducting a pearson's correlation analysis prior to applying PCA on each subset. Even though dimensionality reduction is not a mandatory exercise on small dataset like the one we used, we think that conducting CCA on larger dataset could be an interresting alternative to PCA. \\

From the linear model forecasting, we noticed that the covariates selected from the canonical correlation analysis are more likely to explain the ethereum 
returns according to the RMSE and the MAE Table \ref{tab:rmse_mae}. The canonical correlation analysis can be considered as a serious alternative to principal component analysis for feature selection.\\

We also evaluate the quality of the selection over out sample observations to validate the result of the test sample. The result in the Table \ref{tab:rmse_mae_prediction} shows that the selection based on canonical correlation analysis is better than the one based on principal component analysis. The next step would be to assess the impact of dimensionality reduction on non linear predicators and to integrate the lead-lag relationship between variables. 

\bibliographystyle{IEEEtranN}
\bibliography{bib}
\nocite{*}
\end{document}